\documentstyle{article}

\def\etal{{et al.\ }}

\def\x2{$\chi^{2}$}

\def\asca{{\it ASCA }}
\def\rosat{{\it ROSAT }}

\def\x2{$\chi^{2}$}

\begin{document}

\title{X-ray luminous star-forming galaxies}

\author{A.L. Zezas$^1$ \thanks{aze@star.le.ac.uk},
I. Georgantopoulos$^2$ \thanks{ig@astro.noa.gr} and  M. J. Ward$^3 \thanks{mjw@star.le.ac.uk}$ \\ 
$^1$ Department of Physics and Astronomy, University of Leicester, 
Leicester, LE1 7RH  \\
$^2$ National Observatory of Athens, Lofos Koufou, Palaia Penteli, 
15236, Athens, Greece \\
$^3$ Department of Physics and Astronomy, University of Leicester, 
Leicester, LE1 7RH } 

\date { }
\maketitle

\vspace{-0.5cm}
\begin{abstract}

We present  results from the cross-correlation of the spectroscopic
atlas of Ho \etal (1995) with the \rosat All-Sky Survey Bright Source
Catalogue, in an attempt to understand the 
X-ray emission mechanisms in nearby galaxies. 
 The resulting sample of 45 galaxies consists 
predominantly of  AGN. However,
there are several starforming galaxies spanning a wide range of X-ray
luminosities ($\sim10^{38}$ - $10^{42}\rm{erg\,s^{-1}}$). We
have analyzed \rosat and \asca data for the two most luminous star-forming 
galaxies, 
namely NGC3310 and NGC3690. We find that their 0.1-10 keV 
X-ray spectra can be
fitted by a soft thermal plasma of kT$\sim0.8$ keV and a harder 
 component with kT$\sim10-15$ keV or a 
 power-law with $\Gamma\sim1.6$. 
These are very similar to the spectra of the 
archetypal star-forming galaxies NGC253 and M82. 
\end{abstract}

\newpage

\large

\section{INTRODUCTION}

A large number of faint (B$<23$) galaxies has been found in deep ROSAT surveys
(eg Boyle et al. 1995, Georgantopoulos et al. 1996). 
These galaxies have high X-ray luminosities ($L_{x}>10^{42}$ $\rm erg~s^{-1}$)
and present narrow emission lines in their optical spectra. 
Despite the high probability of chance coincidences 
(ie field galaxies lying accidentally in the error box of the x-ray source), 
 cross-correlations of the X-ray positions with 
 deep optical images have proved the existence of this population   
at a high level of significance (eg Roche et al. 1995, Almaini et al. 1997). 
Schmidt et al. (1998) obtained Keck spectra for 
50 X-ray sources in the Lockman hole. They find that the large  
majority of these host an AGN although some fraction of galaxies
cannot be excluded.  

Now, we have the opportunity to study the 
properties of such luminous galaxies in the local Universe,
using the ROSAT all-sky survey (RASS). 
 Boller et al. (1992) 
 cross-correlated the IRAS Point Source Catalogue with the RASS. 
Preferential spectroscopic follow-up observations of 
the highest X-ray luminosity objects (Moran et al. 1997) 
show  that the vast majority of galaxies 
with $L_{x}>10^{42}$ $\rm erg~s^{-1}$ are AGN.
 Here instead, we have carried out a
cross-correlation of the sample of Ho \etal (1996, 1997) and the \rosat All
Sky Survey Bright Source Catalogue (RASS-BSC).   
 The spectroscopic sample of 
  Ho \etal  contains moderate resolution, 
 high-signal-to-noise spectra of {\it all}
northern galaxies with B$<$12.5 providing a complete
sample of the galactic activity in the nearby universe. The 
important advantage of the Ho et al. sample is the pre-existing, 
 very good quality spectra which give us
unambiguous classifications for all the galaxies. 
The RASS-BSC contains almost
18000 sources found in the  all-sky survey carried out during the
first years of the ROSAT mission. 
 Our aim is to understand the X-ray emission mechanisms in the nearby 
galaxies and to test  whether a class of X-ray
luminous ($L_x\sim 10^{41-42}\,\rm{erg\,s^{-1}}$) star-forming 
 galaxies exists.

\section{THE CROSS-CORRELATION}

 The results of the cross-correlation are presented in table 1. 
  There are 45 coincidences
within 1 arcmin distance from the optical galaxy. On the basis of the 
 sky density of the RASS-BSC sources, we
expect less than 1 to be by chance. Columns 1, 2 and 3 contain the
names and the coordinates of the objects; the X-ray luminosities
calculated using the RASS-BSC count rates and a power-law of $\Gamma=2$ 
 are listed in column 4 (we use an $H_o=65 \rm km~s^{-1}~Mpc^{-1}$); 
finally, in column 5 we give 
the spectroscopic classifications from  Ho \etal (1997). 
 We note that the sample is by no
means statistically complete, due to the non-uniform coverage of
the sky in the RASS. 

 From table 1 we see that the large majority of 
the X-ray galaxies are AGN (Seyferts but also some LINERS).
However, there are seven star-forming galaxies while 
 there is also a large fraction  (6 galaxies) 
of late type or normal galaxies.
 From the seven starforming galaxies of our sample two are
well studied dwarf starforming galaxies ( NGC5204, NGC4449) 
with X-ray luminosities of
$\sim 10^{38}-10^{39}\rm{ergs^{-1}}$ (Della Ceca, Griffiths \& Heckman 1996).
 The archetypal 
star-forming galaxy M82  ($7.5\times10^{40} 
\rm{erg\,s^{-1}}$) is also in our sample. An interesting 
 finding  is the presence of
two star-forming galaxies with luminosities above $10^{41}\rm{ergs^{-1}}$  
reaching the luminosities of low luminosity AGN. 
A peculiar object is NGC5905 which although has a very
high luminosity of $1.5\times10^{42} \rm{erg\,s^{-1}}$ in the RASS, has shown 
a significant decline in X-ray flux in subsequent observations 
(Bade et al. 1996). 
However,  its optical spectrum has not any
signatures of AGN activity.

\section{NGC3690 AND NGC3310}

From the starburst galaxies in this sample we present here results on
the two most luminous ones,  
NGC3310 and NGC3690.  Both are nearby galaxies at
distances of 19.6 Mpc and 63.2 Mpc 
respectively. Both galaxies appear to be in interacting pairs/mergers.  
NGC3690 forms an interacting pair with IC694. Their
separation is 21'' which translates to $\sim6$ kpc at the assumed
distance. For NGC3310 there is also strong evidence that it is the 
remnant of a recent merger, according to anomalies found in its
rotation curve (Mulder \etal 1985), and its disturbed
morphology (Ballick and Heckman, 1981). Evolutionary synthesis 
 modelling of the optical and
infrared spectra of these galaxies has shown that the age of the
starburst is about 10Myr (Pastoriza \etal, 1993 and Nakagawa \etal,
1989 for NGC3310 and NGC3690 respectively). Especially in the case of
the latter Nakagawa \etal find that the two bursts have different
properties, implying either different ages or different Initial Mass
Functions (IMF).

\subsection{X-ray data analysis.}

We have obtained the \rosat (PSPC and HRI) and \asca
observations of these galaxies from the archive. After following the
standard screening procedure, we extracted PSPC and ASCA SIS and GIS spectra
along with HRI images. 
The HRI images show that the soft X-ray emission is extended in NGC3310. 
In NGC3690 the emission comes from three distinct components 
(the two correspond to the nuclei of NGC3690 and IC694) which 
again appear to be spatially resolved. 

We fitted the \rosat and \asca spectra together. The spectral fitting
results are presented in table 2. We found that they are fitted with a
optically thin thermal plasma of temperature $\sim 0.8$keV and either
a hot thermal plasma (kT $\sim10-15$keV), or a power-law with
$\Gamma\sim1.5-1.6$. 
The spectral fits for the soft band are suggestive for a thermal
origin of the X-ray emission, arising from diffuse hot gas, probably 
associated with a galactic super-wind (Heckman \etal 1996). 
 However, the origin of the hard X-rays is
still unclear as there are more than one possible mechanisms which can
produce the observed spectrum. A power-law spectrum can be
produced either by X-ray binaries or Inverse Compton scattering of the
 starburst infrared photons by the supernova generated 
 relativistic electrons, while
hot gas (and X-ray binaries) give a thermal plasma spectrum. These
results are similar to the results found for the prototypical
starburst galaxies M82 and NGC253 (Ptak \etal 1997, Moran and Lehnert
1997), suggesting a common X-ray emission mechanism in star-forming
galaxies spanning a wide range of luminosities. 
 Better signal-to-noise ratio  and higher energy 
X-ray spectra but mainly high resolution X-ray imaging 
are needed in order to draw any conclusions on the origin
of the hard X-ray emission in these galaxies.

\section{CONCLUSIONS}	

 We have presented our results on the cross-correlation of the sample
of Ho \etal (1995) with the RASS-BSC. Our intent was to search for X-ray
luminous star-forming  galaxies and to probe the X-ray content of the
nearby universe. The cross-correlation gives 45 objects 
within a radius of 1 arcmin. Although this sample is
dominated by AGN (mainly Seyferts), there are seven 
star-forming galaxies spanning a large range of X-ray luminosities
($\sim 10^{38}$-$10^{42}\rm{ergs^{-1}}$) rivaling 
the luminosities of low luminosity AGN. 
We have analyzed archival observations of the two most luminous 
star-forming galaxies in our sample namely NGC3690 and NGC3310. 
We find that their soft spectra are fitted by an optically thin thermal
plasma of temperature $\sim0.8\rm{keV}$. 
The hard X-ray emission can be fit either with a high temperature 
thermal plasma ($\rm kT \sim 10-15$ keV) or a flat power-law 
 ($\Gamma\sim 1.6$).  These results 
 are similar to those found for the prototypical
starburst galaxies M82 and NGC253, suggesting a common X-ray emission mechanism in star-forming
galaxies over a large range of luminosities.
 The combination of AXAF 
 and  XMM observations   
will shed more light on the origin of the hard X-ray emission 
which currently remains unknown in all star-forming galaxies.

\section*{References}

Almaini, O., Shanks, T., Griffiths, R.E., Boyle, B.J., Roche, N., Georgantopoulos, I., Stewart, G.C., 1997, MNRAS, 291, 372  \\ 
Bade N. Komossa S. and Dahlem M., 1996, A\&A, 309,35L \\
Ballick B. and Heckman T., 1981, A\&A, 96, 271 \\
Boller \etal, 1992, A\&A, 261, 57 \\
Boyle, B.J., McMahon, R., Wilkes, B.J., Elvis. M., 1995, MNRAS, 276, 315 \\
Della Ceca R., Griffiths R.E., Heckman T.M., 1997, ApJ, 485, 581 \\
Heckman T.M. \etal, 1996, in The enviroment and Evolution of Galaxies,
edited by J.M. Shull and H.A. Thronson, Jr. (Kluwer, Dordrecht) \\
Ho L.C., Filippenko A.V. and Sargent W.L., 1995, ApJS, 98, 477 \\
Ho L.C., Filippenko A.V., and Sargent W.L., 1997, ApJS, 112, 315 \\
Georgantopoulos, I. \etal, 1996, MNRAS, 280, 276  \\
Moran E.C. and Lehnert L.D., 1997, ApJ, 478, 172 \\
Moran E.C., Halpern L.P., Helphand D.J., 1996, ApJS, 106,341 \\
Mulder P.S. and Van Driel W., 1996, 1986, A\&A, 309,403 \\
Nakagawa T., \etal ., 1989, ApJ, 340, 729 \\
Pastoriza \etal, 1993, MNRAS, 260, 177 \\
Ptak \etal, 1997, AJ, 113, 1286 \\
Roche N. \etal, 1995, MNRAS, 273L, 15 \\
Schmidt, M.  \etal 1998, A\&A, 329, 495 \\   

\end{document}